\documentclass[aps, 12pt, prd, onecolumn, tightenlines, notitlepage, superscriptaddress, nofootinbib, preprintnumbers, floatfix,showkeys]{revtex4-1}

\usepackage{tabularx} % extra features for tabular environment
\usepackage{amsmath}  % improve math presentation
\usepackage{graphicx} % takes care of graphic including machinery
\usepackage[margin=1in,letterpaper]{geometry} % decreases margins
\usepackage{subcaption}
\usepackage[frak=esstix]{mathalpha}
\usepackage[final]{hyperref} % adds hyper links inside the generated pdf file
\usepackage{braket}

\hypersetup{
	colorlinks=true,       % false: boxed links; true: colored links
	linkcolor=blue,        % color of internal links
	citecolor=blue,        % color of links to bibliography
	filecolor=magenta,     % color of file links
	urlcolor=blue         
}

\usepackage{hyperref}
\begin{document}
\preprint{IFIC 24/13}
\title{Quantum Decoherence Effects: a complete treatment}
\author{Gabriela Barenboim}\email{gabriela.barenboim@uv.es}
\affiliation{Instituto de Física Corpuscular, CSIC-Universitat de València, C/Catedrático José Beltrán 2, Paterna 46980, Spain}
\affiliation{Departament de Física Teòrica, Universitat de València,
C/ Dr. Moliner 50, Burjassot 46100, Spain}
\author{Alberto M. Gago}\email{agago@pucp.edu.pe}
\affiliation{Sección Física, Departamento de Ciencias, Pontificia Universidad Católica del Perú, Apartado 1761, Lima, Perú}
\date{\today}

\begin{abstract}
Physical systems in real life are inextricably linked to their surroundings and never
completely separated from them. Truly closed systems do not exist. The phenomenon of decoherence, which is brought about by the interaction with the environment, removes the relative phase of quantum states in superposition and makes them incoherent. 
In neutrino physics, decoherence, although extensively studied has only been analyzed  thus far, exclusively in terms of its dissipative characteristics. While it is true that dissipation, or the exponential suppression, eventually is the main observable effect, the exchange of energy between the medium and the system, is an important factor that has been overlooked up until now. In this work, we introduce this term and analyze its consequences.
\end{abstract}

\maketitle

\section{Introduction}

We do not know yet the absolute neutrino masses. We do know however that these masses are non-zero and non-degenerate. We learned that much because we have observed neutrino flavor transitions. There, thanks to the fact that the mass differences are negligible as compared to the energy uncertainty in the generation and detection processes, the different mass eigenstates cannot be identified and therefore, the states of the created and detected neutrinos, which are flavor eigenstates, are expressed as a coherent superposition of mass eigenstates. The dynamics of the neutrinos is governed by the quantum Liouville equation or von Neumann equation for the density matrix. As a result, the evolution is unitary, and coherence—represented by the density matrix's off-diagonal terms—is preserved during the propagation.

This coherence, however, can be lost, as exemplified by solar neutrinos (depending on their production point) during their journey from the interior of the Sun to the Earth. In this scenario, decoherence arises from the separation of the wave packet, corresponding to the neutrino mass eigenstates, over distances due to their disparate group velocities. Nevertheless, this process remains unitary and can be addressed within the standard framework.

%This coherence, however, can be lost, as for example happens with solar neutrinos in their journey from the edge of the Sun to the Earth. The decoherence, in this case, is due to  wave packet dissipation, which results from the separation of neutrino mass states over large distances because of their disparate group velocities. This process is still unitary and can be dealt with the standard framework.

Likewise, when the neutrinos are weakly connected to the environment and get entangled with it during their propagation, a different kind of decoherence can occur. If the environment is added to our Hilbert space, the evolution becomes unitary once more. But because of the vastness of the environment and our ignorance of its properties, we are forced to trace over its degrees of freedom, which gives, in general, rise to non-unitarity. Nevertheless, under the conditions assumed in our formalism, probability will be conserved even in this case.

%Any real physical system is never truly isolated from its environment, but inescapably interacts with it. This interaction causes the phenomenon of decoherence, which eliminates the relative phase of quantum states in superposition so that they can no longer be seen to interfere. 

In neutrino physics, this type of decoherence, called quantum decoherence, has been studied so far, only focusing on its dissipative aspects. The main reason is that decoherence is thought to occur as a byproduct of thermalization, where an open system equilibrates by exchanging energy with its environment. Although it is true that in the long term, dissipation, that is, the exponential suppression,  takes over, exchanging energy with the medium, {\it{i.e.}} extracting or dumping energy into the system, is an aspect that cannot be neglected and has been ignored so far. 

To take into account this exchanging energy with the medium, here we will add to the standard Hamiltonian an interaction term between the neutrino system and the field from the environment, mimicking the well-known atom-field Hamiltonian interaction that produces the Rabi oscillations in the context of quantum optics \cite{Rabi:1936lvg}. Like in quantum optics, this effective interaction term can potentially inject energy into the neutrino system and extract energy from it. The exploration of the phenomenology of this interaction term is well-motivated since it has been largely ignored in the current literature on quantum decoherence and neutrinos.

Furthermore, we assume that gravity is the source of the environment, the so-called gravitational decoherence, and therefore is blind to the particle-antiparticle difference. It is important to note that recently \cite{Oppenheim:2018igd}, it has been pointed out that gravity, even being fundamentally classical, can cause decoherence in a coherent quantum system that is coupled to the diffusion of the metric and its conjugate momenta. The latter reinvigorated the importance of studying how quantum decoherence affects any quantum coherent system, such as the neutrino system.     

Specifically, our approach posits that gravity induces both the field-environment interaction and the associated decoherence effects. This scenario arises when the Effective Average Action (EAA) is applied in the context of quantum gravity \cite{Satz:2010uu, DOdorico:2016poc}, using a bimetric framework with a constant background metric and a quantum fluctuating metric. In this framework, the quantum fluctuations in the metric can lead to decoherence, while the background metric may introduce, for instance, violations of the equivalence principle effects. These effects can be incorporated into the Hamiltonian, being that, under certain conditions, can emulate Rabi oscillations.

\vspace{1cm}
\section{Open quantum system approach}
The evolution of the neutrino system within a pervasive environment is achieved by solving the Lindblad equation \cite{Lindblad:1975ef}, which is used for a non-model-dependent environment:

\begin{equation}
\frac{d\rho(t)}{dt} = -i \left[H,\rho(t)\right] + L\left[\rho(t)\right]
\label{lindblad}
\end{equation}

In this equation, $\rho(t)$ represents the neutrino density matrix over time, and $L\left[\rho(t)\right]$ is the linear map \cite{Benatti:2000ph} that encompasses non-standard damping effects. This term opens the possibility of evolving from pure states to mixed states.

$L\left[\rho(t)\right]$ is defined as follows:

\begin{equation}
L\left[\rho(t)\right] = -\frac{1}{2} \sum_{j} \left\lbrace A^{\dagger}_{j} A_{j} \rho(t) +
\rho(t) A^{\dagger}_{j} A_{j} \right\rbrace
+\sum_{j} A_{j} \rho(t) A^{\dagger}_{j}
\end{equation}

Here, $A_{j}$ represents a set of operators with $j=1,2,...,n^2-1$ for $n$ neutrino generations. These operators must be hermitian to ensure the Rover time increase of the Von Neumann entropy. 

On the other hand, the usual way to deal with Eq.~(\ref{lindblad}), assuming a two-neutrino system, is to write 
the operators $\rho$, $H$ and $A_j$ as: 
\begin{equation}
\rho =\frac{1}{2}\sum{\rho_\mu \sigma_\mu}, \,\, \,\, H =\frac{1}{2}\sum{h_\mu \sigma_\mu},\,\,\,\, A_j=\frac{1}{2}\sum{a^j_\mu \sigma_\mu}
\end{equation} 
where $\mu$ is running from 0 to 3, $\sigma_0$ is the
identity matrix and $\sigma_k$, are the Pauli matrices which obey 
the condition $\left[ \sigma_i,\sigma_j \right]=2 i \epsilon_{ijk} \sigma_k$. Given
the Hermiticity of the $A_j$, the matrix $L_{\mu\nu} (\equiv L[\rho(t)])$ is real and symmetric and it is 
written as: $L_{kj}=\frac{1}{2} \sum_{l,m,n}{(\vec{a}_{n} . \vec{a}_{l}) \epsilon_{knm} \epsilon_{mlj}} $ 
with 
the components $L_{\mu0}=L_{0\mu}=0$ due to the probability conservation. Furthermore, since the elements $L_{kj}$ are defined using scalar products they satisfy the Cauchy-Schwartz inequalities and the complete positivity condition, which is to assure that 
the eigenvalues of $\rho(t)$ are positive at any time. 

Considering all the above, the 
Eq.~(\ref{lindblad}) can be decomposed as: 
\begin{equation}
\dot{\rho}_0=0, \,\,\,\, \dot{\rho}_k=\left( H_{kl}+ L_{kl} \right) \rho_{l} = M_{kl} \rho_{l}
\end{equation}
with $H_{kl}=\sum_{j} h_j \epsilon_{jlk}$. Thus, the solution of $\rho(t)$ given in its matrix form is: 
\begin{equation}
\varrho(t) = e^{\textbf{M}t} \varrho(0)
\end{equation}
where $\varrho$ is a three-dimensional column vector and $\textbf{M}\equiv M_{kl}$. The $e^{\textbf{M}t}$ can be expressed by:
\begin{equation}
e^{\textbf{M}t} = \textbf{D} e^{\textbf{M}_{D} t} \textbf{D}^{-1}
\equiv \left[ e^{\textbf{M}t} \right]_{il}= {\cal{M}}_{il} (t) = \sum_k D_{ik} e^{\lambda_k} D^{-1}_{kl}
\end{equation}
with $\textbf{M}_{D} =\textbf{Diag} \left(\lambda_1,\lambda_2, \lambda_3\right)$. Therefore, we can get the neutrino oscillation probability:
\begin{equation}
P_{\nu_\alpha \rightarrow \nu_\beta} = \text{Tr}\left( \rho^\alpha(t)\rho^\beta \right) 
= \frac{1}{2}+\frac{1}{2} \sum_{i,l} \rho^{\beta}_i(0) 
{\cal{M}}_{il}(t) \, \rho^{\alpha}_l(0)
=\frac{1}{2}+\frac{1}{2} (\varrho^\beta(0))^T\varrho^\alpha(t)
\label{Prob2DDeco}
\end{equation}

\subsection{Hamiltonian of the model}
In our approach the effective (total) Hamiltonian comprises two pieces:
\[\hat{H}_{\text{eff}}=\hat{H}_{\text{osc}}+\hat{H}_{\text{int}},\]
the $\hat{H}_{\text{osc}}$ is the standard Hamiltonian for the two-neutrino oscillation system, which is given by:
\begin{equation}
\hat{H}_{\text{osc}} =
\left( \begin{array}{cc }
\Delta &  \\
  & -\Delta  \end{array} \right)
= \Delta \hat{\sigma_3}
\end{equation}
Here, $\Delta = \Delta m^2/2 E_\nu$. The second piece $\hat{H}_{\text{int}}$
entails the interaction between the neutrino system and the environment-field, and is given by:
\begin{equation}
\hat{H}_{\text{int}} 
=
\left( \begin{array}{cc }
0 &   e^{i \alpha}\lambda \\
 e^{-i\alpha}\lambda  & 0 \end{array} \right)
 = 
\lambda \cos{\alpha} \, \hat{\sigma_1} 
-\lambda \sin{\alpha} \, \hat{\sigma_2}
= \lambda\left[\hat{\sigma}_{+} e^{i \alpha}+\hat{\sigma}_{-} e^{-i \alpha} \right]
\end{equation}
This Hamiltonian term resembles a non-quantized version of the atom-field interaction Hamiltonian used in quantum optics, where $\lambda$ represents the intensity of the neutrino-environment field interaction, while $\alpha$ is a complex phase inherited from the environment field. Meanwhile, the terms $\hat{\sigma}{+}$ and $\hat{\sigma}{-}$ will allow the neutrino conversion and survival modes, being their role in the atom-field interaction to produce the Rabi oscillations, a well-known phenomenon in quantum optics \cite{alicki}.
This term, which violates Lorentz invariance but preserves CPT, was initially noted in \cite{Benatti:2001fa}, yet it has not been thoroughly examined or discussed in the existing literature on quantum decoherence and neutrinos. Its presence distinguishes our study from others. It indeed has some similarities to the well-known Standard Model effective  (SME) coefficients, as we will see later. Here, we assume that the coherent field arises from the environment, having this one a gravitational origin. Thus, the interaction of this field with the neutrino system cannot differentiate between particles and antiparticles. As a result, the environment (or the quantum foam) will play two roles. On one side, it induces decoherence in the neutrino system, while on the other, it generates a coherent field that can interact with this system. 

As a concrete example, these off-diagonal Hamiltonian terms can be generated from equivalence principle violation effects. This can be achieved when the condition $\theta=\theta_G+ \frac{\pi}{4}$ is satisfied, where $\theta$ is the mixing angle connecting the mass eigenstates and flavor eigenstates, while $\theta_G$ is the mixing angle connecting the latter eigenstates with the gravitational ones \cite{Pantaleone:1992ha,Halprin:1995vg}. On the other hand, the decoherence effects, as we mentioned earlier, can arise from stochastic fluctuations from the metric field, which also has a (quantum) gravitational origin \cite{Bernabeu:2006av, Mavromatos:2005bu}. It is good to mention that, based on the formalism developed in our work, it is possible to interpret that the coherent field is independent of the environment. Indeed, the off-diagonal Hamiltonian terms in the 
mass eigenstates basis can also be obtained when a CPT-odd interaction 
term is added and the condition $\theta=\theta_b+ \frac{\pi}{4}$ being $\theta_b$ the mixing angle connecting the flavor basis with the CPT-Odd neutrino interaction eigenstates.

Therefore, in this set-up, $\lambda$ is the dimension-full interaction strength, and $\alpha$ is a  real phase. It has to be emphasized that this interaction does not change sign when going from neutrino to antineutrino, unlike the so-called non-standard interactions. Moreover,  bounds on this coupling do not exist yet.  The known searches for  nonstandard interactions of this type  by measuring the Casimir forces or torsion pendula could potentially bound exclusively a Yukawa-type deviations from Newtonian gravity for the element $\lambda_{ee}$  if $SU(2)_L$ symmetry holds. However, similar bounds on non-diagonal elements do not exist. Although there are no specific limits for $\lambda$, we can expect that its magnitude is in the same order of magnitude as the decoherence parameter, given our assumption of the origin of the interacting field mentioned above. Interestingly, the experimental sensitivity to $\lambda$ for a given experimental scenario is quite similar to that of the decoherence parameter, as we will show later. Therefore, it is not unreasonable to consider the magnitude of $\lambda$ and the decoherence parameters as compatible from the perspective of the experimental power for testing the parameter.

It is also important to notice that $\lambda$ does not need to be constant and can have an arbitrary energy dependence. Different energy dependencies will lead to different distortions, and therefore different phenomenologies. Mass-varying neutrinos~\cite{Fardon:2003eh}, for example,  are just one of the many models that can be obtained.  However, in this work  we want to present a complete framework and not focus on a particular energy dependence, and therefore we will keep $\lambda $ constant. The extension to any energy dependence is straightforward. By keeping the model simple, we can make transparent that there is a strong correlation between distortions to the kinematic phase and the mixing angle and they cannot be arbitrarily tuned.

\vspace{1cm}
\subsection{Neutrino Oscillation Probability} 
\subsubsection{Without Quantum Decoherence Effects}
  Although, according to our hypothesis, the non-diagonal Hamiltonian term in the $\hat{H}_{\text{eff}}$ is a contribution from the environment, in this calculation, we will determine the neutrino oscillation probability without incorporating the dissipative terms. This oscillation probability (without decoherence effects) has a very rich phenomenology on its own and, indeed, constitutes the primary focus of this study.

 For performing the aforementioned calculation we need to diagonalize the $\hat{H}_{\text{eff}}$, which turns out to be:  
\begin{equation}
\hat{H}_{\text{Diag}} =
\frac{1}{2 E_\nu}
\left( \begin{array}{cc }
\Delta m_{\text{eff}}^2&  \\
  & -\Delta m_{\text{eff}}^2 \end{array} \right) = 
   \frac{\Delta m_{\text{eff}}^2}{2 E_\nu} \, \hat{\sigma_3}
\end{equation}
with a new 
effective square mass difference:
\begin{equation}
\Delta_{\text{eff}}=\frac{\Delta m_{\text{eff}}^2}{2 E_\nu} = \sqrt{\displaystyle{\Delta^2 +  \lambda^2 } } \equiv 
\Delta m_{\text{eff}}^2 =
\sqrt{\displaystyle{(\Delta m^2)^2 + 4 \, \lambda^2 \, E_\nu^2 }}
\end{equation}
which in the case $\lambda \ll \Delta $ implies\footnote{It is important to keep in mind that $\Delta \sim  10^{-21} \mbox{GeV}$ for the atmospheric mass difference and energies of a few GeV}
\begin{equation}
 \frac{\Delta m_{\text{eff}}^2}{2 E_\nu} = \Delta\left(1 + \frac{ \lambda^2}{2 \Delta^2} \right)
 =\frac{\Delta m^2}{2 E_\nu} +
 \left (\frac{\lambda^2}{\Delta m^2} \right) E_\nu
\end{equation}
%{\textcolor{red}{the term $\frac{ \lambda^2}{2 \Delta}$ is similar to the $\delta b$ in Barger's paper.}}
It is interesting to note that $\Delta \sim \left( \frac{10^{-21}}{\mbox{GeV}} \right) \mbox{GeV}^2$ for the atmospheric mass squared differences, then, for $\frac{\lambda}{\Delta} \sim 0.1$  and 10 GeV of  neutrino energy will have  $\lambda \sim 10^{-23} \mbox{GeV}$ which is in the order of the sensitivity of  the decoherence parameters. On the other hand, in this limit, a clear parallel can be drawn between the $\left(\frac{ \lambda^2}{\Delta m^2}\right)$ term and the SME $c$-coefficients \cite{Kostelecky:2003fs}, which appear alongside a kinematic phase scaling linearly with $E_\nu$. Even upon further expansion of $\frac{\Delta m_{\text{eff}}^2}{2 E_\nu}$, the correspondence between our model and the $c$-coefficients associated with (odd) higher powers of $E_\nu^{2n-1}(n=2,..)$ remains valid, with the $c$-coefficients being proportional to $\left(\frac{ \lambda^{2n}}{(\Delta m^2)^{2n-1}}\right)$. From this perspective, our model offers a framework to establish a direct relationship between the $c$-coefficients of higher-order terms in the SME and the higher-order contributions in the series expansion of $\frac{\Delta m_{\text{eff}}^2}{2 E_\nu}$.

The another ingredient for the diagonalization (and the calculation of the neutrino oscillation probability) is the mixing matrix, which connects the neutrino flavor states with the effective vacuum mass eigenstates. This is given by:
\begin{equation}
  U(\beta+\theta,\alpha)= \left(
\begin{array}{cc}
 e^{i (\pi -\alpha)}  \cos (\beta+\theta)  & \sin (\beta+\theta)  \\
 e^{-i\alpha} \sin (\beta+\theta)  & \cos (\beta+\theta) \\
\end{array}
\right)
\label{mixingfull}
\end{equation}

Where:

\begin{equation}
\tan^2(\beta+\theta) = \dfrac{\lambda  \sin 2 \theta + \sqrt{\displaystyle{\Delta^2 +  \lambda^2  } }-\Delta \cos 2\theta}{-\lambda \sin 2 \theta + \sqrt{\displaystyle{\Delta^2 +  \lambda^2  } }+\Delta \cos 2\theta}
\end{equation}
here, $\theta$ is the vacuum mixing angle. This formula  exhibits the resonant behavior we see in the case of non-standard interactions, when
\begin{equation}
 \lambda = \Delta \cot 2\theta 
 \label{resonant}
\end{equation}

However in our case, the resonance will be present in both channels.
The mixing angle in a medium, then  will increase  if
\[  \lambda \sin 2 \theta  > \Delta \cos 2\theta \]
or
\[ 2  \lambda E  > \Delta m^2 \cot 2\theta \]
and will decrease in the opposite case.

In the small interaction regime,  

\begin{eqnarray}
 \tan^2(\beta+\theta) &=&   \tan ^2\theta +\frac{\lambda  \, \tan\theta \, \sec
   ^2\theta }{\Delta} \nonumber \\
   \tan^2(\beta+\theta) &=& \tan ^2\theta \left( 1 + \frac{2 \lambda E}{\sin \theta \, \Delta m^2} \right)
\end{eqnarray}

%Given that as both, the mass difference and the mixing angle are shifted the same way for neutrinos and antineutrinos  in our gravitational pervasive environment, the propagation in such a medium does not induce a fake, medium induced,  CP violation and therefore not only CPT is conserved. Therefore, it can be anticipated that: 

Given that both the mass difference and the mixing angle are similarly affected for neutrinos and antineutrinos in our gravitationally pervasive environment, their propagation in such a medium does not induce a spurious, medium-induced CP violation. Therefore, not only CPT is conserved but also CP\footnote{For genuine CPT violation in the mass terms see \cite{Barenboim:2002tz}}, consequently, we can anticipate the following CPT conserving relations: 
\[ P(\nu_\alpha \longrightarrow \nu_\alpha)= P(\overline{\nu}_\alpha \longrightarrow \overline{\nu}_\alpha)\]
\[ P(\nu_\alpha \longrightarrow \nu_\beta)= P(\overline{\nu}_\beta \longrightarrow \overline{\nu}_\alpha)\]
and also the ones that preserves CP (in the two generation case)
\[ P(\nu_\alpha \longrightarrow \nu_\beta)= P(\overline{\nu}_\alpha \longrightarrow \overline{\nu}_\beta)\]
In the three generation case, (genuine) CP violation will appear, and this equation will be no longer valid. Medium effects will induce fake CP violation as well. The inclusion of (non-diagonal) decoherence effects from the environment can reveal the presence of Majorana CP phases in the neutrino oscillation formula~\cite{Benatti:1997xt, Carrasco, Capolupo}.

%Aside from our interaction's inability to differentiate between neutrinos and antineutrinos, similar results can be obtained from Reference \cite{Barger}, (i.e., by setting $2 \theta_b = \pi/2 + 2\theta_m$, where both mixing angles are as defined in this reference, with $\theta_m = \theta$).

Now, with all the ingredients in hand, we can give the expression of the neutrino oscillation 
probability~\footnote{The neutrino oscillation probability can also be calculated 
using the density matrix formalism (Eq.~(\ref{Prob2DDeco})). 
The $\rho_e \left(0\right)$ 
and $\rho_\mu \left(0\right)$ are given in 
the \hyperref[sec:AppendixA]{Appendix A}}:
\begin{equation}
P_{\nu_\alpha \rightarrow \nu_\beta} =\sin^2 2(\beta+\theta) \sin^2 \left(\frac{\sqrt{\Delta^2 + \lambda^2}}{2}L\right)
\label{Prob2dModel}
\end{equation}
for $\alpha \neq \beta $. As expected, the formula above is similar, at some level, to the vacuum Rabi Oscillations. Particularly, the $\sqrt{\Delta^2 + \lambda^2}$ plays the same role as the Rabi frequency, setting the pace for energy exchange between the neutrino system and the environment-field 
with $\Delta$ be analogous to the detuning. While the oscillation amplitude 
is given by: 
\begin{equation}
\sin^2 2(\beta+\theta) = 
\dfrac{\left(\Delta \sin 2\theta +\lambda \cos 2\theta \right)^2 }{\Delta^2 +\lambda^2}
\label{sin2tot}
\end{equation}
the oscillation amplitude clearly confirms the resonant condition $\lambda = \Delta \cot 2\theta$ quoted before. It is straightforward to note that the oscillation amplitude also satisfies an anti-resonant condition at $\lambda=-\Delta \tan{2\theta}$.

Now, considering the oscillation 
amplitude and the Eq.~(\ref{Prob2dModel}) in
\begin{figure}
    \centering
    \includegraphics{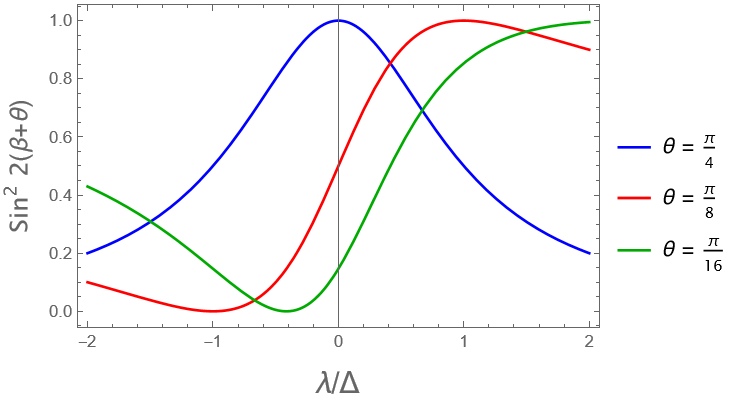}
    \caption{Amplitude of the oscillation as a function of $\lambda / \Delta $ for different choices of the vacuum mixing angle}
    \label{fig:amposc}
\end{figure}
the small interaction regime (i.e. 
expanding the probability up to the first order in $\lambda/ \Delta$) we get the
following expression for the two-generation probability for the model: 
\begin{eqnarray}
P_{\nu_\alpha \rightarrow \nu_\beta} 
& \approx &
\sin^2 2\theta
\left( 1 + \left(\frac{4 E_\nu \lambda}
{\Delta m^2}   \right)\cot{2 \theta} \right)
\sin^2 \left(\frac{\Delta}{2}L\right) \\ \nonumber \\
& \approx &
\left( \sin^2 2\theta + \left(\frac{2 E_\nu \lambda}
{\Delta m^2}   \right)\sin{4 \theta} \right)
\sin^2 \left(\frac{\Delta}{2}L\right)
\label{Probapprox2dModel}
\end{eqnarray}
%%%%
from the above expression, we can see that the amplitude of the oscillation can be amplified or reduced, depending on the mixing angle. Specially interesting is the case 
$\theta \approx \pi/4 $, where the suppression is remarkable, as we will see later. 
Notice that this modulation of the mixing angle, which scales linearly with the energy,
and unlike to matter effect does not depend on the density of the medium, offers a unique
opportunity to search/bound these kinds of effects. It should be noted that if the SME CPT preserving but Lorentz violating terms are the first-order contribution of a full decoherence model, its modification of the kinematic phase should be accompanied by this higher-order modification of the mixing angle. 

Considering the full expression of $\sin^2 2(\beta+\theta)$ as given in Eq.~(\ref{sin2tot}), we illustrate its behavior in Figure(\ref{fig:amposc}) for three different values of $\theta$: $\pi/4, \pi/8$ and $\pi/16$. For $\theta=\pi/4$, the peak for $\sin^2 2(\beta+\theta)$ is achieved when $\lambda=0$ as ruled by the resonant behavior given in Eq.~(\ref{resonant}), while, for the rest of the values of $\lambda/\Delta$, it is suppressed regardless of whether $\lambda$ is positive or negative  This suppression is due to the attenuation factor $(1+(\lambda/\Delta)^{2})^{-1}=(1+(2E_\nu \lambda/\Delta m^2)^{2})^{-1}$, which diminishes as $|\lambda/\Delta|$  increases (or the increase of $E_\nu$ for a fixed $\lambda$), as we previously observed in Eq.~(\ref{Probapprox2dModel}). In the scenario where $\theta=\pi/8$, the peak, as expected from Eq.~(\ref{resonant}), is at $\lambda=\Delta$, while 
the valley is, in general, at $\lambda=-\Delta \tan{2\theta}$ which in this case is at $\lambda=-\Delta$. On the other hand, the suppression is governed by the factor $(1+2(\lambda/\Delta)/(1+(\lambda/\Delta)^{2}))$. For $|\lambda/\Delta|< 1$, this factor approximately takes the form $(1+2(\lambda/\Delta))$. This explains why within the range $[-1,1]$, the $\sin^2 2(\beta+\theta)$ increases (decreases) for positive (negative) $\lambda$. Outside this range, the amplitude of $\sin^2 2(\beta+\theta)$ diminishes as the attenuation factor converges to $(1+2/(1+(\lambda/\Delta)^{2}))$. For $\theta=\pi/16$, although we observe a behavior akin to that of $\theta=\pi/8$, it is different the interval in which $\sin^2 2(\beta+\theta)$ either increases or decreases due to the distinct weighting of $\lambda$ and $\Delta$ in the complete expression of $\sin^2 2(\beta+\theta)$. The peak is reached at $ \lambda= 2.4142\,\Delta$ while the valley is at $\lambda=-0.4142\,\Delta$.
 
\begin{figure}
    \centering
    \includegraphics{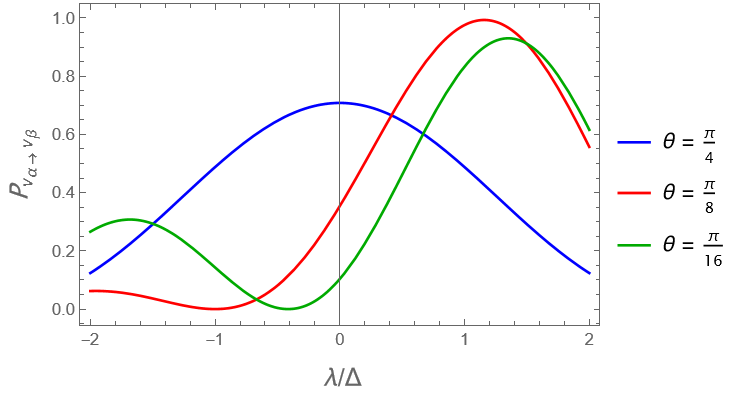}
    \caption{Transition probability as a function of $\lambda / \Delta $ for different choices of the vacuum mixing angle for $\Delta m^2 L /(4 E) \approx 1$}
    \label{transitions}
\end{figure}

For completeness, the probability transitions are shown in Figure(\ref{transitions}) where the effects of the mixing angle distortions due to the interaction with the environment are moderated by the oscillation driven by the kinematic phase, which in turn is also modified as shown before. There it can be clearly seen that the amplitude of the oscillation, is the one  depicted in Figure(\ref{fig:amposc}), while the position of the maximum of the oscillation is dictated by the modulation of the effective mass difference.

\subsubsection{With Quantum Decoherence Effects}
Given the non-diagonal nature of our Hamiltonian, $\hat{H}_{\text{eff}}$, we find ourselves in a scenario akin to the one shown in \cite{Carpio:2017nui}. Thus, the most straightforward approach is where the decoherence matrix remains unaltered after rotating the non-diagonal basis to the diagonal one. The latter property, as highlighted in \cite{Carpio:2017nui}, allows us to directly engage with $\hat{H}_{\text{eff}}$ and 
to compute the neutrino oscillation probability, including the quantum decoherence effects, following the standard methodology \cite{Gago:2002na,Barenboim:2004wu}. The decoherence matrix that satisfies this criterion is $\textbf{L}=-\textbf{Diag} \left(\gamma,a, \gamma \right)$ \cite{Carpio:2017nui}. 

Using the Eq.~(\ref{Prob2DDeco}) we can write the neutrino oscillation probability (the corresponding ${\cal{M}}_{il}$ 
are listed on the Appendix B): 
\begin{eqnarray}
P_{\nu_{\alpha}\rightarrow \nu_{\beta}}(t) = \frac{1}{2}&+&\frac{1}{2}\left\{\left[\left( \rho_{1}^{\alpha}\rho_{1}^{\beta}+\rho_{2}^{\alpha}\rho_{2}^{\beta}\right) \cosh {\left(\frac{\Omega t}{2}\right)}  \right. \right. \nonumber \\
&+&  \left. \left(\frac{ 2\Delta_{\text{eff}} \left(\rho_{1}^{\alpha}\rho_{2}^{\beta}-\rho_{2}^{\alpha}\rho_{1}^{\beta}\right)+\left(-\gamma+a \right) \left(\rho_{1}^{\alpha}\rho_{1}^{\beta}-\rho_{2}^{\alpha}\rho_{2}^{\beta} \right)}{\Omega} \right) \sinh{\left(\frac{\Omega t}{2} \right)}
\right]e^{-\left(\gamma+a\right)t}  \nonumber \\
&+& \rho_{3}^{\alpha}\rho_{3}^{\beta} e^{-2\gamma t} \Bigg\}\,,
\label{final}
\end{eqnarray}
where $\Omega = \sqrt{-4 \Delta_{\text{eff}}^2 +\left(\gamma-a \right)^2}$. 
Considering $\rho_{1}^\alpha =2 \,\mathfrak{Re} (U_{\alpha 1}^{\ast}U_{\alpha 2})$, 
$\rho_{2}^\alpha =-2 \,\mathfrak{Im} (U_{\alpha 1}^{\ast}U_{\alpha 2})$, and 
$\rho_{3}^\alpha = |U_{\alpha 1}|^2 -|U_{\beta 2}|^2$ we have:
\begin{eqnarray}
\rho_{1}^{\alpha}\rho_{1}^{\beta}+\rho_{2}^{\alpha}\rho_{2}^{\beta} &=& 
4 \,\mathfrak{Re} \left( U^{*}_{\alpha 1}
   U_{\alpha 2}
   U_{\beta 1}
   U^{*}_{\beta 2}
  \right) \nonumber\\ 
\rho_{1}^{\alpha}\rho_{2}^{\beta}-\rho_{2}^{\alpha}\rho_{1}^{\beta} &=& 
4 \,\mathfrak{Im} \left( U^{*}_{\alpha 1}
   U_{\alpha 2}
   U_{\beta 1}
   U^{*}_{\beta 2}
  \right) \nonumber\\ 
\rho_{3}^\alpha \rho_{3}^\beta &=& -1 + 2 \sum_{l=1,2}
 |U_{\alpha l}|^2 
  |U_{\beta l}|^2 =-1 + 2\delta_{\alpha \beta}
  - 4 \,\mathfrak{Re} \left( U^{*}_{\alpha 1}
   U_{\alpha 2}
   U_{\beta 1}
   U^{*}_{\beta 2}
  \right) 
\end{eqnarray}
With the relations above we can re-write Eq.~(\ref{final}) as follows:
\begin{eqnarray}
P_{\nu_{\alpha}\rightarrow \nu_{\beta}}(t) &= & \frac{1}{2} + \frac{1}{2}\left\{\left[4 \,\mathfrak{Re} \left( U^{*}_{\alpha 1}
   U_{\alpha 2}
   U_{\beta 1}
   U^{*}_{\beta 2}
  \right) \cosh {\left(\frac{\Omega t}{2}\right)}  \right. \right. \nonumber \\
&+&  \left. \left( 
\frac{8 \,\Delta_{\text{eff}} \mathfrak{Im} \left( U^{*}_{\alpha 1}
   U_{\alpha 2}
   U_{\beta 1}
   U^{*}_{\beta 2}
  \right)
+
4 \,\left(-\gamma+a \right) \mathfrak{Re} \left( U^{*}_{\alpha 1}
   U_{\alpha 2}
   U^{*}_{\beta 1}
   U_{\beta 2}
  \right)}{\Omega} \right) \sinh{\left(\frac{\Omega t}{2} \right)}
\right]e^{-\left(\gamma+a\right)t}  \nonumber \\
&+& \left( -1 + 2\delta_{\alpha \beta}
  - 4 \,\mathfrak{Re} \left( U^{*}_{\alpha 1}
   U_{\alpha 2}
   U_{\beta 1}
   U^{*}_{\beta 2}
  \right)\right) e^{-2\gamma t} \Bigg\}\,,
\label{final2}
\end{eqnarray}
Replacing the mixing matrix elements from Eq.~(\ref{mixingfull}) we got the 
following expression:
\begin{eqnarray}
P_{\nu_{\alpha}\rightarrow \nu_{\beta}}(t) = \frac{1}{2}
&-&\frac{1}{2}e^{-2\gamma t} 
+\left( \delta_{\alpha \beta} +\frac{1}{2}\sin^2 2(\beta+\theta) \right)\, e^{-2\gamma t}
\nonumber \\
&-& \frac{1}{2}\sin^2 2(\beta+\theta) \cosh \left(\frac{\Omega t}{2}\right) 
e^{-\left(\gamma+a\right)t} 
\nonumber \\
&+& \frac{1}{2} \left(\gamma -a \right) \sin^2 2(\beta+\theta) \cos 2\alpha 
\left( \frac{\sinh{\left(\frac{\Omega t}{2} \right)}}{\Omega} \right) 
e^{-\left(\gamma+a\right)t} 
\label{final3}
\end{eqnarray}
 substituting $|\Omega|\rightarrow 2\Delta_{\text{eff}}$, $\gamma=a=0$ in the latter equation 
the Eq.~(\ref{Prob2dModel}) can be easily recovered. 

\begin{figure}
    \centering
    \includegraphics{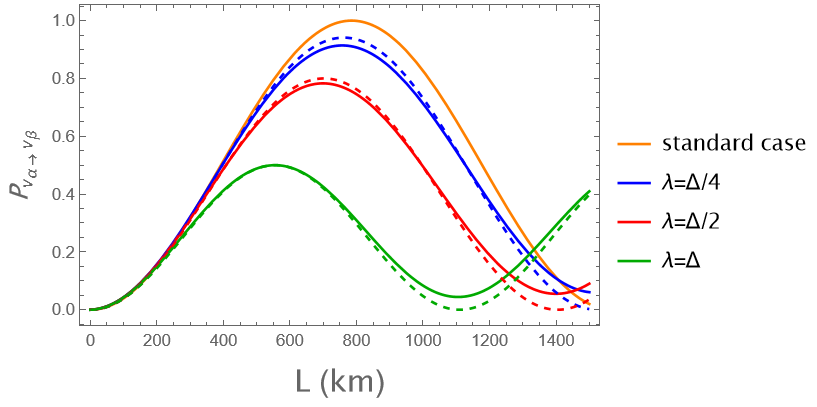}
    \caption{Transition probabilities, for a maximal mixing angle in vacuum, are presented in three different scenarios, $\lambda/\Delta=\left\{ 1/4,1/2,1\right\}$, with (solid) and without (dashed) exponential suppression. The oscillation has been set to reach its maximum  for $L\simeq 800$Km. The standard scenario is also shown for comparison.     In the scenarios, including exponential suppression,$\gamma=8.2 \cdot 10^{-24} {\mbox{GeV}}= \left(24000 {\mbox{km}}\right)^{-1} $ has been used.}
    \label{comparison}
\end{figure}

\begin{figure}
    \centering
    \includegraphics{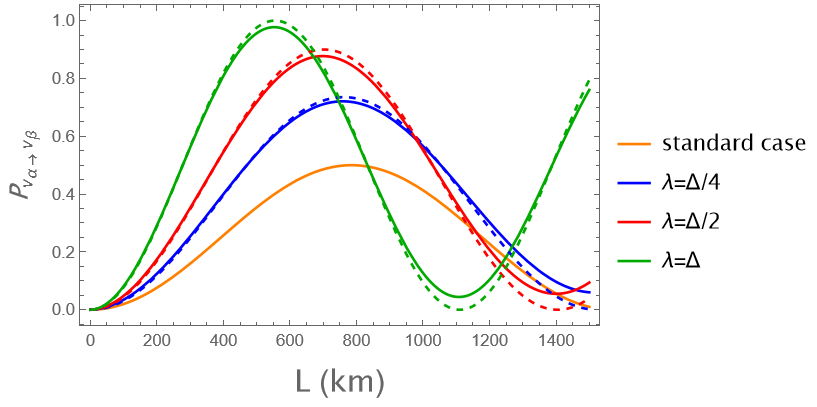}
    \caption{Same plot as in Figure(\ref{comparison}) but for $\theta =\pi/8$}
    \label{comparison2}
\end{figure}

To stress how the new (complete) decoherence scenario differs from what has been done so far, we can analyze a very simplified scenario. Lets concentrate on the transition probability only, {\it i.e.} $\alpha \neq \beta $ and assume  $\gamma=a$. In this case,
the last term vanishes, $\Omega\rightarrow 2 i \Delta_{\text{eff}}$ and the probability takes the following form
\begin{eqnarray}
P_{\nu_{\alpha}\rightarrow \nu_{\beta}}(t) = \frac{1}{2}
&-&\frac{1}{2}\left( 1- 2 \sin^2 2(\beta+\theta) \sin^2\left( 
\frac{\Delta m_{\text{eff}}^2  L}{4 E_\nu}\right)\right)\, e^{-2\gamma t}\nonumber\\
= \frac{1}{2}
&-&\frac{1}{2}\left( 1- \dfrac{2\left(\Delta \sin 2\theta +\lambda \cos 2\theta \right)^2 }{\Delta^2 +\lambda^2}\sin^2\left( 
\frac{\sqrt{\Delta^2 + \lambda^2}}{2}L\right)\right)\, e^{-2\gamma t}
\label{simpli}
\end{eqnarray}
%where it is pretty transparent that the energy exchange with the medium is parametrized by $\lambda=-\Delta \tan{2\theta}$ and is independent of the $\gamma$-effect. 
%Figure(\ref{comparison}) represents the transition probabilities in three different scenarios where the new term is present, with (solid) and without (dashed) $\gamma$-term, for a maximal mixing angle in vacuum, where $\sin(\Delta L/2)$ has been set to one for $L\simeq 800$Km. No matter effects have been included. For the sake of comparison, the no decoherence scenario is also shown. Figure(\ref{comparison2}) describes the same scenarios for a mixing angle of $\theta=\pi/8$.
where it is pretty transparent that the intensity of the neutrino-environment field interaction parametrized by $\lambda=-\Delta \tan{2\theta}$ is independent of the $\gamma$-effect, being that the frequency of the energy exchange is controlled by $\sqrt{\Delta^2 + \lambda^2}/2$. Figure(\ref{comparison}) represents the transition probabilities in three different scenarios where the new term is present, with (solid) and without (dashed) $\gamma$-term, for a maximal mixing angle in vacuum $\theta=\pi/4$, where $\sin(\Delta L/2)$ has been set to one for $L\simeq 800$ km. No matter effects have been included. For the sake of comparison, the no decoherence scenario is also shown. As expected, and given that $\theta=\pi/4$, we can see that as long as $\lambda$ increases, the oscillation amplitude decreases in relation to the standard case. Similarly, the shift of the oscillation frequency (energy exchange frequency) is larger as $\lambda$ grows. Figure(\ref{comparison2}) describes the same scenarios for a mixing angle of $\theta=\pi/8$. For this case, the effect is the opposite for the oscillation amplitude, as it increases when $\lambda$ increases. The behavior of the shift in oscillation frequency is similar to the previous case; i.e., the shift is larger for larger $\lambda$.

For the sake of clarity, we have not incorporated in the discussion the matter effects which will be present if the neutrinos propagate in a medium. Such effects can be trivially incorporated into the formulation by the simple replacement of the mass difference and mixing angles in vacuum by the corresponding ones in a medium.

\section{Conclusions}

In this work, we go beyond other neutrino decoherence studies performed so far, which have only considered the dissipative effects of the environment. We introduce the possibility of interaction between the neutrino system and an environmental field. This Hamiltonian term, mimicking the atom-field Hamiltonian used in quantum optics, represents an energy exchange between the neutrino system and the environmental field—losing or gaining energy, akin to Rabi oscillations. This unexplored interaction opens up a very interesting phenomenology.

One of our most interesting findings is that this interaction can induce resonant behavior, maximizing oscillation amplitude. Conversely, the oscillation amplitude can be suppressed. The amplification or suppression of the oscillation amplitude, resulting from the oversight of this interaction, could be a potential source of distortion in the measurement of oscillation parameters for future neutrino accelerator experiments, such as those referenced in \cite{Arguelles:2022tki}. Specially interesting are the prospects at DUNE and T2HK. 

Besides, these effects can be probed in the light of the actual high-energy atmospheric neutrino data, or we can further assess their impact in the high-energy astrophysical neutrino observations. The reference to these test scenarios is based on the fact that the resonant condition is achieved when $2 E_\nu \lambda = \Delta m^2 \cot{2\theta}$, implying that, with increasing neutrino energy, we can be sensitive to smaller values of $\lambda$.

On the other hand, the interference phase $\Delta m_{\text{eff}}^2/4 E_\nu =\sqrt{\Delta^2 + \lambda^2}/2$, when considering a small $\lambda$, could be hard to probe at ${\mathcal{O}}(1 \,\text{GeV})$   energies. This is due to the shift at the first order scaling as $\propto \lambda^2 E_\nu$. Therefore, utilizing the high-energy atmospheric data set for this purpose would be advantageous. It is evident that there is still much to be done, such as expanding our study to the three-neutrino generation framework, which includes matter effects. This expansion also involves, as mentioned earlier, contrasting the neutrino-environment field interaction in its three-neutrino generation version with current data and considering future projections.

%has not been explored so far and leads to completely new and rich phenomenology which can let the neutrino system to exchange energy from the environment-field, losing or gaining it. 

All in all, there isn't a single quantum system that is  completely closed or isolated. Rather, coherent dynamics—as characterized by a Schrödinger equation—usually is only  valid for very short timescales before the coupling between the open system and its surroundings takes over, resulting in coherence loss and the beginning of more classical behavior. Therefore, specially in the case of neutrinos, which travel long distances and come almost unaltered from the past, a complete understanding of the phenomenology associated to the interaction with the environment not restricted only to decoherence is mandatory.

\section*{Acknowledgments}

\noindent
This work is supported by the Spanish grants  CIPROM/2021/054 (Generalitat Valenciana) and PID2020-113775GB-I00 (AEI/10.13039/501100011033), by the European ITN project HIDDeN (H2020-MSCA-ITN-2019/860881-HIDDeN), 
Marie Skłodowska-Curie Staff Exchange grant  ASYMMETRY (HORIZON-MSCA-2021-SE-01/101086085-ASYMMETRY) and by
{\it{Dirección
de Fomento de la Investigación}} at Pontificia Universidad
Católica del Perú, through Grants No. DFI-2021-0758 and CONCYTEC through Grant
No.60-2015-FONDECYT. A.M. Gago wants to express his gratitude to the Departament de Física Teòrica, Universitat de València for their warm hospitality during the development of this work. He also wants to thank F. de Zela and E. Massoni for the valuable discussions and suggestions.

\renewcommand{\theequation}{A-\arabic{equation}}
\setcounter{equation}{0}  % reset counter 
\section*{Appendix A:\, $\rho_e \left(0\right)$ and $\rho_\mu \left(0\right)$}
\label{sec:AppendixA} 
\begin{equation}
\rho_e \left(0\right)= \left(
\begin{array}{cc}
\cos^2 (\beta+\theta) & - e^{i \alpha} \cos (\beta+\theta) \sin (\beta+\theta) \\

    e^{-i \alpha}\cos (\beta+\theta) \sin (\beta+\theta) & \sin^2 (\beta+\theta) \\
    \end{array}
    \right)
    \end{equation}
\begin{equation}
\rho_\mu \left(0\right)=
\left(
\begin{array}{cc}
\sin^2 (\beta+\theta) & e^{i \alpha} \cos (\beta+\theta) \sin (\beta+\theta) \\
e^{-i \alpha}\cos (\beta+\theta) \sin (\beta+\theta) & \cos^2 (\beta+\theta) \\
\end{array}
\right)
\end{equation}

\renewcommand{\theequation}{B-\arabic{equation}}
\setcounter{equation}{0}  % reset counter 
\section*{Appendix B:\, ${\cal{M}}_{il}$ elements}

\begin{eqnarray}
{\cal{M}}_{11}(t) &=& \left(\cosh {\left(\frac{\Omega t}{2}\right)} + \left(-\gamma+a \right) 
\frac{\sinh{\left(\frac{\Omega t}{2} \right)}}{\Omega}
\right) e^{-\left(\gamma+a\right)t} \nonumber \\ 
{\cal{M}}_{12}(t) &=& 2 \Delta_{\text{eff}} \, \frac{\sinh{\left(\frac{\Omega t}{2} \right)}}{\Omega} 
\nonumber \\ 
{\cal{M}}_{21}(t) &=&-2 \Delta_{\text{eff}} \, \frac{\sinh{\left(\frac{\Omega t}{2} \right)}}{\Omega} 
\nonumber \\
{\cal{M}}_{22}(t) &=& \left(\cosh {\left(\frac{\Omega t}{2}\right)} - \left(-\gamma+a \right) 
\frac{\sinh{\left(\frac{\Omega t}{2} \right)}}{\Omega}
\right) e^{-\left(\gamma+a\right)t} 
\end{eqnarray}
where ${\cal{M}}_{13}(t) ={\cal{M}}_{31}(t) = 0$, 
${\cal{M}}_{23}(t) ={\cal{M}}_{32}(t) = 0$ 
and $\Omega=\sqrt{-4 \Delta_{\text{eff}}^2 +\left(\gamma-a \right)^2}$


\begin{thebibliography}{99}
%\cite{Rabi:1936lvg}
\bibitem{Rabi:1936lvg}
I.~I.~Rabi,
%``On the Process of Space Quantization,''
Phys. Rev. \textbf{49} (1936) no.4, 324
doi:10.1103/PhysRev.49.324 ; 
%132 citations counted in INSPIRE as of 17 Jan 2024
%\cite{Rabi:1937dgo}
%\bibitem{Rabi:1937dgo}
I.~I.~Rabi,
%``Space Quantization in a Gyrating Magnetic Field,''
Phys. Rev. \textbf{51} (1937) no.8, 652
doi:10.1103/PhysRev.51.652 
%257 citations counted in INSPIRE as of 17 Jan 2024


%\cite{Oppenheim:2018igd}
\bibitem{Oppenheim:2018igd}
J.~Oppenheim,
%``A Postquantum Theory of Classical Gravity?,''
Phys. Rev. X \textbf{13} (2023) no.4, 041040
doi:10.1103/PhysRevX.13.041040
[arXiv:1811.03116 [hep-th]].
%50 citations counted in INSPIRE as of 19 Dec 2023

%\cite{Satz:2010uu}
\bibitem{Satz:2010uu}
A.~Satz, A.~Codello and F.~D.~Mazzitelli,
%``Low energy Quantum Gravity from the Effective Average Action,''
Phys. Rev. D \textbf{82} (2010), 084011
doi:10.1103/PhysRevD.82.084011
[arXiv:1006.3808 [hep-th]].
%38 citations counted in INSPIRE as of 09 Jul 2024

%\cite{DOdorico:2016poc}
\bibitem{DOdorico:2016poc}
G.~D'Odorico, A.~Codello and C.~Pagani,
%``The Background Effective Average Action Approach to Quantum Gravity,''
Springer Proc. Phys. \textbf{170} (2016), 233-239
doi:10.1007/978-3-319-20046-0\_27
%0 citations counted in INSPIRE as of 09 Jul 2024


%\cite{Lindblad:1975ef}
\bibitem{Lindblad:1975ef}
G.~Lindblad,
%``On the Generators of Quantum Dynamical Semigroups,''
Commun. Math. Phys. \textbf{48} (1976), 119
doi:10.1007/BF01608499
%1997 citations counted in INSPIRE as of 19 Dec 2023 


%\cite{Benatti:2000ph}
\bibitem{Benatti:2000ph}
F.~Benatti and R.~Floreanini,
%``Open system approach to neutrino oscillations,''
JHEP \textbf{02} (2000), 032
doi:10.1088/1126-6708/2000/02/032
[arXiv:hep-ph/0002221 [hep-ph]].
%106 citations counted in INSPIRE as of 19 Dec 2023

\bibitem{alicki}
 R.~Alicki and K.~Lendi, Lect. Notes Phys. \textbf{286} (1987), Springer-
Verlag, Berlin.


%\cite{Benatti:2001fa}
\bibitem{Benatti:2001fa}
F.~Benatti and R.~Floreanini,
%``Massless neutrino oscillations,''
Phys. Rev. D \textbf{64} (2001), 085015
doi:10.1103/PhysRevD.64.085015
[arXiv:hep-ph/0105303 [hep-ph]].
%88 citations counted in INSPIRE as of 08 Jan 2024

%\cite{Pantaleone:1992ha}
\bibitem{Pantaleone:1992ha}
J.~T.~Pantaleone, A.~Halprin and C.~N.~Leung,
%``Neutrino mixing due to a violation of the equivalence principle,''
Phys. Rev. D \textbf{47} (1993), R4199-R4202
doi:10.1103/PhysRevD.47.R4199
[arXiv:hep-ph/9211214 [hep-ph]].
%84 citations counted in INSPIRE as of 09 Jul 2024

%\cite{Halprin:1995vg}
\bibitem{Halprin:1995vg}
A.~Halprin, C.~N.~Leung and J.~T.~Pantaleone,
%``A Possible violation of the equivalence principle by neutrinos,''
Phys. Rev. D \textbf{53} (1996), 5365-5376
doi:10.1103/PhysRevD.53.5365
[arXiv:hep-ph/9512220 [hep-ph]].
%96 citations counted in INSPIRE as of 09 Jul 2024

%\cite{Bernabeu:2006av}
\bibitem{Bernabeu:2006av}
J.~Bernabeu, N.~E.~Mavromatos and S.~Sarkar,
%``Decoherence induced CPT violation and entangled neutral mesons,''
Phys. Rev. D \textbf{74} (2006), 045014
doi:10.1103/PhysRevD.74.045014
[arXiv:hep-th/0606137 [hep-th]].
%58 citations counted in INSPIRE as of 09 Jul 2024

%\cite{Mavromatos:2005bu}
\bibitem{Mavromatos:2005bu}
N.~Mavromatos and S.~Sarkar,
%``Liouville decoherence in a model of flavor oscillations in the presence of dark energy,''
Phys. Rev. D \textbf{72} (2005), 065016
doi:10.1103/PhysRevD.72.065016
[arXiv:hep-th/0506242 [hep-th]].
%41 citations counted in INSPIRE as of 09 Jul 2024

%\cite{Fardon:2003eh}
\bibitem{Fardon:2003eh}
R.~Fardon, A.~E.~Nelson and N.~Weiner,
%``Dark energy from mass varying neutrinos,''
JCAP \textbf{10}, 005 (2004)
doi:10.1088/1475-7516/2004/10/005
[arXiv:astro-ph/0309800 [astro-ph]].
%424 citations counted in INSPIRE as of 01 Apr 2024

%\cite{Kostelecky:2003fs}
\bibitem{Kostelecky:2003fs}
V.~A.~Kostelecky,
%``Gravity, Lorentz violation, and the standard model,''
Phys. Rev. D \textbf{69} (2004), 105009
doi:10.1103/PhysRevD.69.105009
[arXiv:hep-th/0312310 [hep-th]].
%1327 citations counted in INSPIRE as of 19 Dec 2023

%\cite{Benatti:1997xt}
\bibitem{Benatti:1997xt}
F.~Benatti and R.~Floreanini,
%``Completely positive dynamics of correlated neutral kaons,''
Nucl. Phys. B \textbf{511} (1998), 550-576
doi:10.1016/S0550-3213(97)00705-0
[arXiv:hep-ph/9711240 [hep-ph]].
%62 citations counted in INSPIRE as of 19 Dec 2023

%\cite{Carrasco-Martinez:2020mlg}
\bibitem{Carrasco}
J.~C.~Carrasco-Mart\'\i{}nez, F.~N.~D\'\i{}az and A.~M.~Gago,
%``Uncovering the Majorana nature through a precision measurement of the CP phase,''
Phys. Rev. D \textbf{105} (2022) no.3, 035010
doi:10.1103/PhysRevD.105.035010
[arXiv:2011.01254 [hep-ph]].
%7 citations counted in INSPIRE as of 08 Jan 2024

%\cite{Capolupo:2020hqm}
\bibitem{Capolupo}
A.~Capolupo, S.~M.~Giampaolo, G.~Lambiase and A.~Quaranta,
%``Discerning the Nature of Neutrinos: Decoherence and Geometric Phases,''
Universe \textbf{6} (2020) no.11, 207
doi:10.3390/universe6110207
%10 citations counted in INSPIRE as of 08 Jan 2024


%\cite{Barenboim:2002tz}
\bibitem{Barenboim:2002tz}
G.~Barenboim and J.~D.~Lykken,
%``A Model of CPT Violation for Neutrinos,''
Phys. Lett. B \textbf{554} (2003), 73-80
doi:10.1016/S0370-2693(02)03262-8
[arXiv:hep-ph/0210411 [hep-ph]].
%94 citations counted in INSPIRE as of 19 Dec 2023

%\cite{Barger:2000iv}
\bibitem{Barger:2000iv}
V.~D.~Barger, S.~Pakvasa, T.~J.~Weiler and K.~Whisnant,
%``CPT odd resonances in neutrino oscillations,''
Phys. Rev. Lett. \textbf{85} (2000), 5055-5058
doi:10.1103/PhysRevLett.85.5055
[arXiv:hep-ph/0005197 [hep-ph]].
%158 citations counted in INSPIRE as of 04 Jul 2024


%\cite{Barger:1999bg}
%\bibitem{Barger}
%V.~D.~Barger, J.~G.~Learned, P.~Lipari, M.~Lusignoli, S.~Pakvasa and T.~J.~Weiler,
%``Neutrino decay and atmospheric neutrinos,''
%Phys. Lett. B \textbf{462} (1999), 109-114
%doi:10.1016/S0370-2693(99)00887-4
%[arXiv:hep-ph/9907421 [hep-ph]].
%181 citations counted in INSPIRE as of 04 Jan 2024


%\cite{Carpio:2017nui}
\bibitem{Carpio:2017nui}
J.~Carpio, E.~Massoni and A.~M.~Gago,
%``Revisiting quantum decoherence for neutrino oscillations in matter with constant density,''
Phys. Rev. D \textbf{97} (2018) no.11, 115017
doi:10.1103/PhysRevD.97.115017
[arXiv:1711.03680 [hep-ph]].
%30 citations counted in INSPIRE as of 19 Dec 2023

%\cite{Gago:2002na}
\bibitem{Gago:2002na}
A.~M.~Gago, E.~M.~Santos, W.~J.~C.~Teves and R.~Zukanovich Funchal,
%``A Study on quantum decoherence phenomena with three generations of neutrinos,''
[arXiv:hep-ph/0208166 [hep-ph]].
%54 citations counted in INSPIRE as of 19 Dec 2023

%\cite{Barenboim:2004wu}
\bibitem{Barenboim:2004wu}
G.~Barenboim and N.~E.~Mavromatos,
%``CPT violating decoherence and LSND: A Possible window to Planck scale physics,''
JHEP \textbf{01} (2005), 034
doi:10.1088/1126-6708/2005/01/034
[arXiv:hep-ph/0404014 [hep-ph]].
%91 citations counted in INSPIRE as of 19 Dec 2023

%\cite{Arguelles:2022tki}
\bibitem{Arguelles:2022tki}
C.~A.~Arg\"uelles, G.~Barenboim, M.~Bustamante, P.~Coloma, P.~B.~Denton, I.~Esteban, Y.~Farzan, E.~F.~Mart\'\i{}nez, D.~V.~Forero and A.~M.~Gago, \textit{et al.}
%``Snowmass white paper: beyond the standard model effects on neutrino flavor: Submitted to the proceedings of the US community study on the future of particle physics (Snowmass 2021),''
Eur. Phys. J. C \textbf{83} (2023) no.1, 15
doi:10.1140/epjc/s10052-022-11049-7
[arXiv:2203.10811 [hep-ph]].
%59 citations counted in INSPIRE as of 18 Jan 2024


\end{thebibliography}
\end{document}